# Photonic band-structure and optical S-matrix analyses of the proposed superconducting nanostructures

Ali Sobhani Khakestar[1], Alireza Kokabi[1], Khashayar Mehrany[1], Mehdi fardmanesh[1]
*Department of Electrical Engineering, Sharif University of Technology, Tehran, Iran*

Special optical structures composed of superconducting nanostructure based on the elemental type-II and layered high-$T_c$ superconductors are proposed. The photonic band structure and the optical S-matrix are calculated for these proposed structures for different geometrical values. Such structures produce terahertz (THz) photonic band gap, which is controlled by the temperature and geometrical parameters. In addition, the proposed structures show controllable optical reflectance and transmittance in the considered terahertz range. The calculations are performed in a special frequency range in which the refractive index of the layered superconductors has negative value according to the previous analyses.

## I. INTRODUCTION

Recently, nanosuperconductivity, the investigation of the physical properties of the superconductor nanostructures, attracted many superconductivity and nanotechnology researchers [1-6]. Superconductivity gap, $E_g$, and critical temperature, $T_c$, are calculated according to the sample geometrical parameters in the range of the nanometer [7]. The transition temperature in the order of 40K for 10nm-width nanowires made of $YBa_2Cu_3O_{7-\delta}$ high temperature superconductor is reported [8-9]. Superconducting transition is also observed in the metallic nanoparticles [10-12]. Such superconducting nanostructures might make it possible to fabricate very high-density superconductor-based devices like as high-resolution arrays of sensitive infrared detectors. Nevertheless, the superconductivity is predicted in the doped graphane, which is a very thin layer and is almost a 2D nanostructure, recently [13].

As intuitively expected, experimental works express that shrinking the superconductor dimensions would decrease the superconductivity gap and shift the critical temperature to the lower values. In other word, the observable effects of the superconductivity phenomenon switch to lower temperatures when the material dimension is reduced below coherence length. This agrees with the mean-field theories, which predict the existence of the superconductivity in a many body systems in contrast with few-atom systems. However, as we mentioned, experimental works show that the superconductivity exists nevertheless in the nanometer scales.

On the other hand, Superconductors, as an adjustable dispersive and nonlinear media with significant anisotropy has been regarded as one of the promising candidates for the next generation of optical and microwave devices [14-16]. The superconductors are also temperature dependent optical mediums, which might be a promising capability in some applications. Consequently, many theoretical and experimental analyses have been performed on the superconductor optical and photonic properties and considerable and applicable effects have been predicted or observed in this area. The possibility of superconductor-based photonic crystals with tunable bandgap is proposed theoretically based on the pancake and Josephson vortex 2D lattices [17-18]. Terahertz radiation in the layered superconductors with the power and the wavelength controllability by some easily adjustable parameters such as magnetic field and electrical biasing is another interested topic, which is concerned a lot in recent years [19]. Superconductors are also widely applied as optical detection sensors such as hot electron, tunnel-junction and hot spot optical, infrared and microwave detectors. Similarities between Josephson plasma waves in layered superconductors and nonlinear optics is also investigated previously [20]. Very recently, negative refractive index, a very much-interested optical property, is also theoretically predicted in the layered superconductors [21].

On the other hand, the optical properties of the nanoparticle networks and other periodic nanostructures are also issue of a large number of investigations. Nanostructures are applied as gratings, photonic crystals and optical absorbers. In addition, tapering structures are also used and investigated as impedance matching method to adjust the waveguide boundaries or cavity-waveguide couplings. According to the above discussion and considering the optical application capability of both superconductors and nanostructures, the nanosuperconductivity might also become a subject of future optical researches.

Motivated by this prediction, here, we wish to draw attention to the feasibility of applying superconductor nanostructures as controllable optical mediums, and we investigate the optical effects of periodic nanostructures based on the layered and type-II superconductors. Thus, two specific nanostructures are proposed to be optical devices. The first one is a network of superconductor nanoparticles and the second one is a periodic stack of layered superconductors. In both structures, the size of superconducting parts is in the range that superconductivity exists based on the experimental works. Thus, these parts are assumed to be in superconducting phase. Such structures might be applied in the superconductor-based optical devices as impedance matching part of boundaries and couplers to enhance some technologically limitation of these devices such as low terahertz radiation power in the stack of intrinsic Josephson junctions. Here, we also calculated the photonic band structure and the optical reflectance of the proposed structured. We show that the proposed structures have some specific optical properties. As it is also shown in the next section, the proposed structures exhibit large temperature dependencies in the considered optical parameters. Here, the effect of geometrical parameters on the corresponding bandgap is also studied.

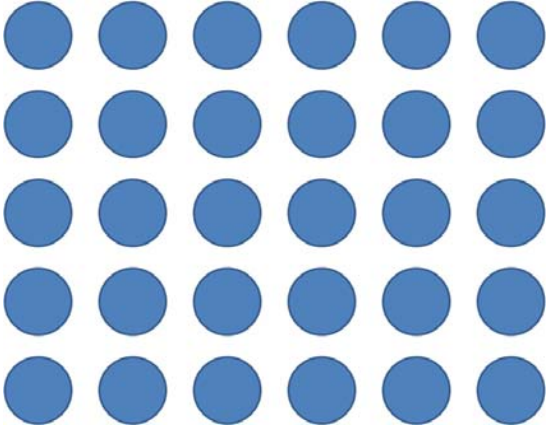

Fig. 1. The structure of nanoparticle network

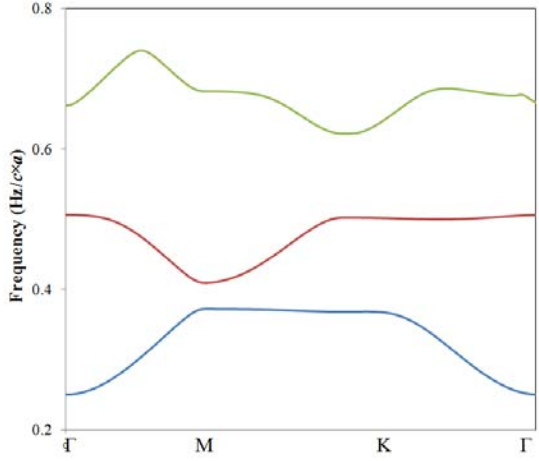

Fig. 2. Photonic band structure of nanoparticle network

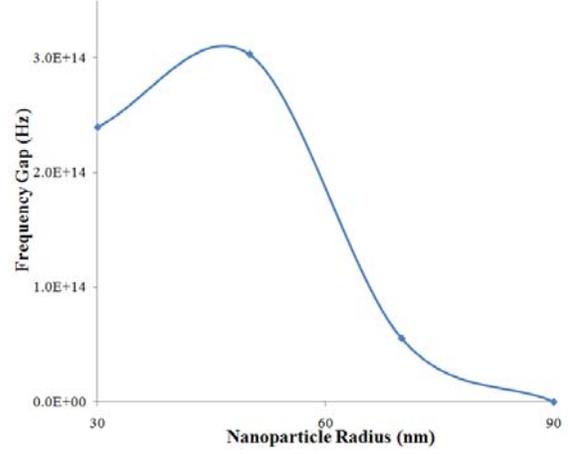

Fig. 3. First photonic bandgap versus nanoparticle radius

## II. TYPE-II BASED STRUCTURE

a. Model Equations

We start from the Drude model of metals which predicts the high frequency conductivity to be in the form of [22]

$$\sigma(\omega) = \frac{\sigma_0}{1 - i\omega\tau}, \quad (1)$$

where $\sigma_0$ is the DC conductivity, $\tau$ is the electron mean-free time and $\omega$ is the angular frequency. $\sigma_0$ is obtained from $\sigma_0 = ne^2\tau/m$ where $e$ is the carrier electric charge, $n$ is the carrier density and m is the carrier effective mass.

Then, according to the two-fluid model of the superconductors, two kinds of carriers contribute in the conductivity below the critical temperature, $T_c$. Thus, the total conductivity is sum of quasiparticles and paired electrons conductivities,

$$\sigma(\omega) = \sigma_n(\omega) + \sigma_s(\omega). \quad (2)$$

According to the Drude model, either of the quasiparticle conductivity, $\sigma_n$, and paired electron conductivity, $\sigma_s$, can be written as

$$\sigma_n(\omega) = \frac{n_n e^2}{m(\gamma - i\omega)}, \quad (3)$$

$$\sigma_s(\omega) = i\frac{n_s e^2}{m\omega}. \quad (4)$$

Here, $n_n$ and $n_s$ are quasiparticle and paired electron densities respectively, $\gamma = 1/\tau$ and the effect of loss damping term in the conductivity of paired electrons is eliminated.

By substitution of $\sigma(\omega)$ in the Maxwell equation,

$$\nabla \times \mathbf{H} = \mathbf{J} + \partial \mathbf{D}/\partial t, \quad (5)$$

and using the equation $\mathbf{J} = \sigma(\omega)\mathbf{E}$, the corresponding effective complex permittivity based on the two-fluid model and Drude model is obtained to be [23-24]

$$\varepsilon_{\text{eff}} = \varepsilon - \frac{i\gamma n_n e^2}{m\omega(\gamma^2 + \omega^2)} + \frac{n_n e^2}{m(\gamma^2 + \omega^2)} + \frac{n_s e^2}{m\omega^2}. \quad (6)$$

Here, $\varepsilon$ is the permittivity of the superconductor media. In the above equation, $n_s$ and $n_n$ are temperature dependent parameters that vary through the following equations:

$$n_s = 1 - \left(T/T_c\right)^2. \quad (7)$$

b. Results and Analyses

We first consider the nanostructures of the elemental type-II superconductors. In type-II superconductors, the coherent length, $\xi$ is in the order of tens of nanometers. These range of values is much smaller than the ones for the type-I superconductors. Thus, the superconductivity might be possible in smaller dimensions as it is confirmed by the experiments [].

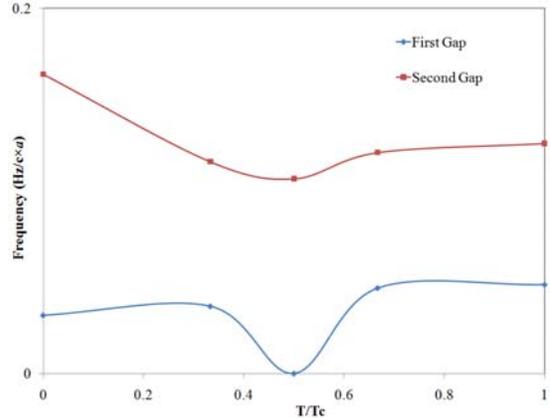

Fig. 4. First and second photonic bandgap versus temperature

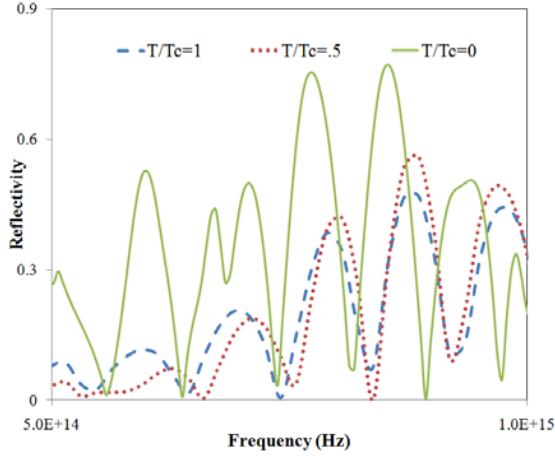

Fig. 5. The reflectivity of nanoparticle network

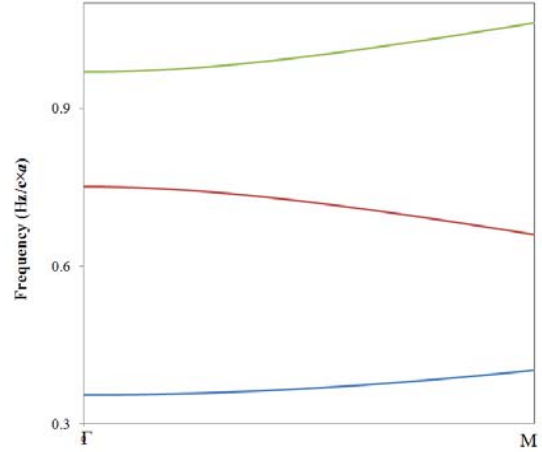

Fig. 7. The band structure of superconductor grating

In the first structure, we assume a network of elemental superconductor nanoparticles. The nanoparticle lattice is supposed to be cubic structure with the nanoparticle radius of $r$ and the center-to-center distance of $d$ as it is shown in the Fig. 1. We choose the values of $d$=200nm and $r$ in the order of 80nm in our analyses.

In the Fig. 2, the typical TE band structure for such a nanoparticle network is shown. As it is obvious from this figure, a gap between the first and the second band and also between the second and third band are opened possibly because of the periodicity of the considered structure. The calculations are done by assuming $r$=70nm and $d$=200nm Fig. 3 presents the variation of first gap with respect to the radius of the superconducting nanoparticles, one of the geometrical parameters. As illustrated, increasing the particle size, initially, enhance the bandgap, while above particle radius of about 50nm, the bandgap begins to decrease. As previously mentioned, the effective permittivity of the superconductors is a temperature dependent parameter. Thus, one expects the bandgap also changes when the temperature is varied. In Fig. 4, such a dependency of both of the bandgaps to temperature is investigated. As it is obvious, the first gap is almost temperature independent except at the half of critical temperature in which the gap is totally vanished.

Finally, we investigate the reflectivity of the nanoparticle network in the range of terahertz. Fig. 5 presents the result of these analyses. The figure shows that in some frequencies the variation of the temperature converts the nanoparticle network from nearly reflective to nearly transparent optical media. It is also concluded from the figure that increasing the temperature decreases the overall reflectivity.

## III. LAYERED SUPERCONDUCTOR-BASED STRUCTURE

a. Model Equations

In contrast to elemental type-II superconductors, in the case of the layered superconductors, the anisotropy of the media induce different values of the permittivity normal and parallel to $c$-axis of the material; [21]

$$\varepsilon_\perp \approx 1 - \varepsilon \frac{s\omega_p^2}{d\omega^2} + \frac{i\sigma_i s}{\omega d}, \tag{8}$$

$$\varepsilon_\parallel \approx \varepsilon\left(1 - \gamma^2 \frac{\omega_p^2}{\omega^2}\right) + \frac{i\sigma_s}{\omega}, \tag{9}$$

where $\omega_p = c/(\lambda_\perp \varepsilon^{1/2})$ denotes the plasma frequency of the intrinsic Josephson junctions in the layered superconductors, $s$ and $d$ are the superconducting and insulating layer thicknesses respectively, $\gamma = \lambda_\perp/\lambda_\parallel$ is the index of the anisotropy, $\sigma_s$ is the quasiparticle conductivity of superconducting layers and $\sigma_i$ is conductivity of the insulating layers. In this equation, $\omega_p, \gamma$ are superconducting temperature dependent parameters. In the above equations, the $\varepsilon_\parallel$ is negative if the frequency be in the range of

$$\sqrt{\frac{s\varepsilon + d}{d}} < \frac{\omega}{\omega_p} < \gamma\sqrt{\frac{\varepsilon(s+d)}{d\varepsilon + s}}, \tag{10}$$

Based on the London theory, the penetration depth is related to the carrier density through the following equation:

$$\lambda = \sqrt{\frac{m}{\mu n_s e^2}}, \tag{11}$$

where $\mu$ is the magnetic susceptibility of the superconducting material. Since $n_s$ is a temperature dependent parameter, thus we expect $\lambda$ to be also a temperature dependent one.

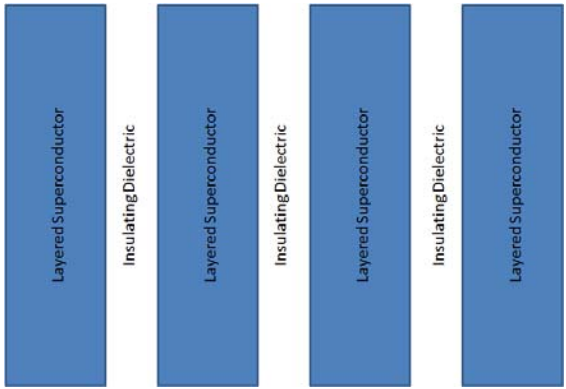

Fig. 6. The structure of superconductor grating

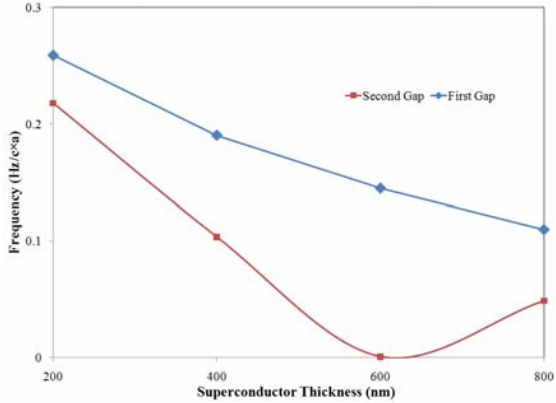

Fig. 8. The first and second bandgap versus superconductor thickness for nanoparticle network

b. Results and Analyses

The second considered structure is a periodic layered grating composed of insulating dielectric and layered superconductors that can be assumed as artificial stack of Josephson junctions or natural anisotropic structures such as BSCCO superconductor. This periodic structure is introduced in the Fig. 6. As we previously mentioned, layered superconductors show many attractive optical properties such as terahertz radiation [19], microwave detection [25] and negative refractive index [21]. Thus, it might be interesting to investigate a periodic structure based on this type of superconductors hoping that such a structure to be applicable in superconductor-based optical devices. Fig. 7 shows the typical TE band structure of such gratings with the assumption of superconductor thickness, $s$=200nm and dielectric thickness, $t$=1800nm. In this figure, the first band lies in the frequency range that superconductor has negative permittivity and the second band is at the frequency range of positive superconductor permittivity. Thus, at the gap between these two bands, the permittivity of the layered superconductor changes sign. This property might have some special applications. The dependency of the first and the second gap to the superconductor thickness is shown in the Fig. 8. At all over the investigated range of superconductor thickness, the first gap is larger than the second one that might be associated to the change of permittivity sign in the first gap. Temperature variation of these two gaps is also studied in the Fig. 9

Fig 10 presents the reflectivity of this superconductor-based 1D grating versus frequency at three different temperatures. Similar to the previous structure, at some frequencies the reflectivity changes about 75% by variation of the temperature. In contrast with that structure, the overall reflectivity is not significantly varied as the temperature changes.

## IV. CONCLUSION

Optical properties of two proposed superconducting nanostructures are investigated. Temperature and dimension dependent photonic bandgaps in the range of the terahertz are observed in such structures. Photonic band gap shows temperature dependent behavior, which could be associated to the variation of the permittivity to the temperature. The terahertz reflectivity of these structures is also extensively varied by temperature and can be adjusted from reflectivity to transparency by temperature variation. In the case of the layered structure, the permittivity changes sign in the first gap and thus for the first and second band the sign of permittivity differs.

## REFERENCES

[1] I. Sochnikov, A. Shaulov, Y. Yeshurun, G. Logvenov and I. Božović, "Large oscillations of the magnetoresistance in nanopatterned high-temperature superconducting films," *Nature Nanotechnol.*, vol. 5, pp. 516–519, Jun. 2010.
[2] M. Tian, J. Wang, N. Kumar, T. Han, Y. Kobayashi, Y. Liu, T. E. Mallouk, and M. H. W. Chan, "Observation of Superconductivity in Granular Bi Nanowires Fabricated by Electrodeposition," *Nano Lett.*, vol. 6, pp. 2773-2780, 2006.
[3] I. W. Chung, S. J. Kwon, S. J. Kim, E. S. Jang, S. J. Hwang and J. H. Choy, "Evidence of Two-Dimensional Superconductivity in the Single Crystalline Nanohybrid of Organic-Bismuth Cuprate," *J. Phys. Chem. B,* vol. 110, No. 33, Jul. 2006.
[4] W. M. Chen, S. S. Jiang, Y. C. Guo, L. Y. Li, K. Shen and S. X. Dou, "Structure and superconductivity of the Y4Ba8Cu12O27 nanosuperconductor," , *J. Supercond.*, vol. 11, pp. 743-748, 1998.
[5] G. Zhang, X. Lu, T. Zhang, J. Qu, W. Wang, X. Li and S. Yu, "Microstructure and superconductivity of highly ordered YBa2Cu3O7−δ nanowire arrays," *Nanotechnol.,* vol. 17, pp. 4252–4256, Aug. 2006.
[6] W. M. Chen, "Nanosuperconductor YBa2Cu3Oy," *Phys. Rev. B,* vol. 57, pp. 7503-7505, Apr. 1998.
[7] G. Brammertz, A. Golubov, A. Peacock, P. Verhoeve, D. J. Goldie and R. Venn, "Modelling the energy gap in transition metal/aluminium bilayers," *Physica C: Superconductivity*, vol. 350, pp. 227-236, Feb. 2001.
[8] K. Xu and J. R. Heath, "Long, Highly-Ordered High-Temperature Superconductor Nanowire Arrays," *Nano Lett.,* vol. 8, pp. 3845-3849, 2008.
[9] S. H. Lai, Y. C. Hsu and M. D. Lan, "Synthesis of Bi2Sr2CaCu2Oy nanowire and its superconductivity," *Solid State Communications*, vol. 148, pp. 452-454, Sep. 2008.
[10] S. Neeleshwar, Y. Y. Chen, C. R. Wang, M. N. Ou and P. H. Huang, "Superconductivity in aluminum nanoparticles," *Physica C: Superconductivity*, vol. 408–410 pp. 209–210, 2004.
[11] F. Y. Wu, C. C. Yang, C. M. Wu, C. W. Wang and W. H. Li "Superconductivity in zero-dimensional indium

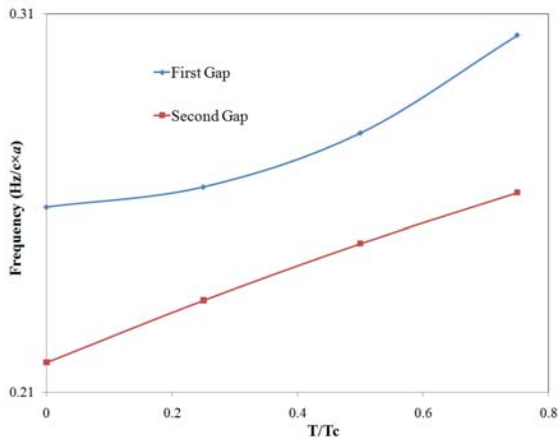

Fig. 9. The first and second bandgap versus superconductor thickness for 1D layered structure

nanoparticles," *J. Appl. Phys.*, vol. 101, pp. 09G111-1-09G111-3, May. 2007.
[12] S. Li, T. White, J. Plevert and C. Q. Sun, "Superconductivity of nano-crystalline MgB2," *Supercond. Sci. Technol.,* vol. 17, pp. S589–S594, Aug. 2004.
[13] G. Savini1, A. C. Ferrari, and F. Giustino, "First-Principles Prediction of Doped Graphane as a High-Temperature Electron-Phonon Superconductor," *Phys. Rev. Lett.,* vol. 105, pp. 037002-1 - 037002-4, 2010.
[14] G. M. Coutts, R. R. Mansour and S. K. Chaudhuri, "High-temperature superconducting nonlinear transmission lines," *IEEE Trans. Microwave Theory Techniques,* vol. 48, pp. 2511 – 2518, Dec. 2000.
[15] V. A. Yampol'skii1, S. Savel'ev1, A. L. Rakhmanov, and F. Nori, "Nonlinear electrodynamics in layered superconductors ," *Phys. Rev. B*, vol. 78, pp. 024511-1 - 024511-9, Jul. 2008.
[16] S. Savel'ev, V. A. Yampol'skii, A. L. Rakhmanov and F. Nori, "Layered superconductors as nonlinear waveguides for terahertz waves," *Phys. Rev. B,* vol. 75, pp. 184503-1-184503-8, May 2007.
[17] H. Takeda and K. Yoshino, "Properties of Abrikosov lattices as photonic crystals," *Phys. Rev. B*, vol. 70, pp. 085109-1 - 085109-5, Aug. 2004.
[18] S. Savel'ev, A. L. Rakhmanov and F. Nori, "Using Josephson Vortex Lattices to Control Terahertz Radiation: Tunable Transparency and Terahertz Photonic Crystals," *Phys. Rev. Lett.,* vol. 94, pp. 157004-1 - 157004-4  Apr. 2005.
[19] M. H. Bae, H. J. Lee and J. H. Choi, "Josephson-Vortex-Flow Terahertz Emission in Layered High-Tc Superconducting Single Crystals," *Phys. Rev. Lett.*, vol. 98, pp. 027002-1 - 027002-4, Jan. 2007.
[20] S. Savel'ev, A. L. Rakhmanov, V. A. Yampol'skii and F. Nori, "Analogues of nonlinear optics using terahertz Josephson plasma waves in layered superconductors," *Nature Phys. 2,* pp. 521 – 525, 2006.
[21] A. L. Rakhmanov, V. A. Yampol'skii, J. A. Fan, Federico Capasso and F. Nori, "Layered superconductors as negative-refractive-index metamaterials," *Phys. Rev. B*, vol. 81, pp. 075101-1 - 075101-6, Feb. 2010.
[22] Introduction to Solid State Physics, C. Kittel, Wiley; 8 edition (November 11, 2004) ISBN-10: 9780471415268
[23] H. Zandi, A. Kokabi, A. Jafarpour, S. Khorasani, M. Fardmanesh and A. Adibi, "Photonic band structure of isotropic and anisotropic Abrikosov lattices in superconductors," *Physica C: Superconductivity,* vol. 467, pp. 51-58, Sep. 2007.
[24] A. Kokabi, H. Zandi, S. Khorasani and M. Fardmanesh, "Precision photonic band structure calculation of Abrikosov periodic lattice in type-II superconductors," *Physica C: Superconductivity,* vol. 460-462, pp. 1222-1223, Sep. 2007.
[25] T. Takashi, U. Takashi and Y. Yoshizumi, "BSCCO intrinsic Josephson junctions for microwave detection," *IEEE transactions on applied superconductivity*, vol. 13, pp. 901-903, 2003.

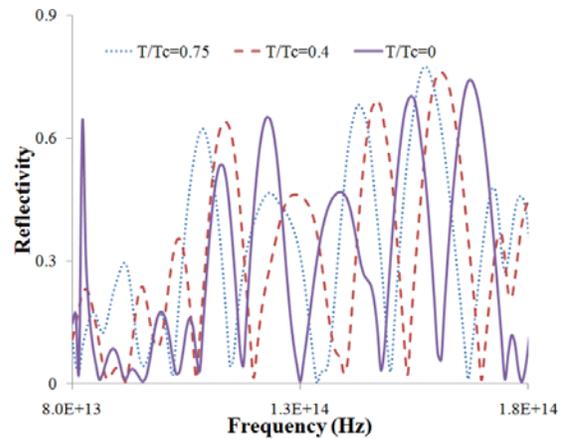

Fig. 10. The reflectivity of 1D layered structure